# Подавление шумов в рентгеновской Фурье-голографии при использовании двухблочного интерферометра из френелевских зонных пластин с общей оптической осью

Л. А. Арутюнян

*Ереванский государственный университет, Ереван, Армения*
*levon.har@gmail.com*

(2 Мая, 2018 г.)

В ранее представленной схеме рентгеновской Фурье-голографии, основанной на двухблочном интерферометре из френелевских зонных пластин с общей оптической осью, налагаются жесткие требования к размерам исследуемого образца. Не соблюдение этих условий приводит к образованию шума в восстановленном изображении. В представленной работе исследуется механизм образования этого шума и возможность его подавления.

## 1. Введение

В работе [1] представлена схема Фурье-голографии для жесткого рентгеновского излучения, основанного на двухблочном интерферометре из френелевских зонных пластин (ФЗП). Показано, что при регистрации неоднородностей образца с низкими пространственными частотами интерферометр работает в режиме деления амплитуды, с равными длинами траекторий в обоих плечах интерферометра. В случае же неоднородностей с высокими частотами интерферометр перестает работать в вышеуказанном режиме и ужесточаются условия, налагаемые на характеристики когерентности исходного излучения.

Наряду с аналитическим исследовании представленной схемы, в отмеченной работе проведена численное моделирование записи голограммы и последующее восстановление изображения предмета. С целью увеличения размеров исследуемого объекта, при 2D моделировании не били соблюдены требования предъявляемым к параметрам интерферометра, необходимые для блокировки других – «нежелательных» каналов распространения излучения. В результате, на нижней половине и внизу восстановленного изображения появились так называемые «интерференционные шумы».

Целю настоящей работы является исследование механизма возникновения этих шумов и возможности их устранения.

## 2. Схема интерферометра и механизм образования шумов

Подробная схема интерферометра для регистрации голограмм представлена на рис.1. Он состоит из двух ФЗП, с общей главной оптической осью, удаленных друг от друга на расстояние $2F$ ($F$ – фокусное расстояние первого порядка дифракции ФЗП). Предметная плоскость находится вправо от второго ФЗП, на расстоянии $F$, а детектор голограммы – за предметной плоскости, на таком же расстоянии от последнего. Непосредственно перед первым ФЗП расположен нож с горизонтальным краем, закрывающий более половины зонной пластины. В предметной плоскости находится непрозрачный экран, с квадратным окном для предмета и круглой диафрагмой с центром на оптической оси, для опорной волны. В плоскости голограммы введена координатная система $(\tilde{x}, \tilde{y})$ с началом координат в точке пересечения оптической оси с голограммой. Ось $\tilde{x}$ направлен по горизонтали, а $\tilde{y}$ – верх по вертикали. Аналогичная координатная система – $(x, y)$, с началом координат на центре восстановленного изображения введена в плоскости изображения.

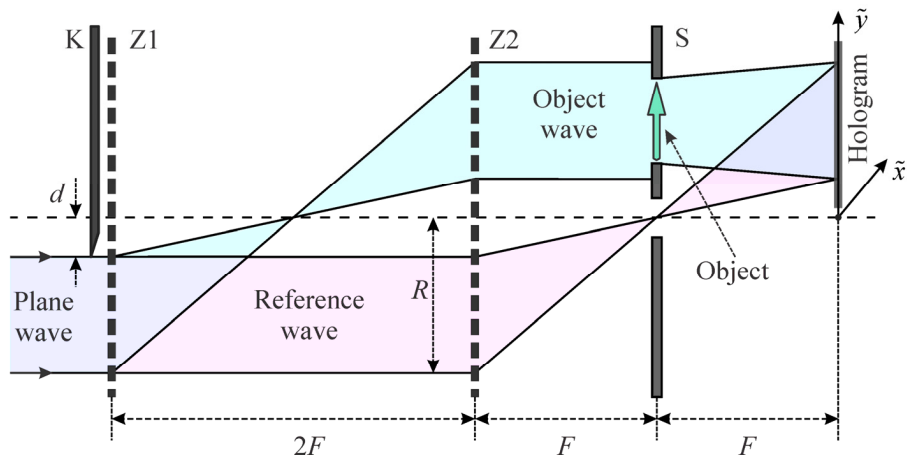

Рис.1. Схема интерферометра для записи голограммы. Z1 и Z2 – первый и второй ФЗП соответственно, K – нож, S – экран в предметной плоскости.

Предметная волна формируется дифракциями первого порядка на обоих ФЗП. Образованная при этом параллельный пучок, проходя сквозь исследуемый объект падает на голограмму. Опорная волна формируется волной, проходящей без дифракции (нулевой порядок дифракции) от первой ФЗП и дифрагированной в первом порядке на втором. Фокусируясь после этого в предметной плоскости и переходя сквозь диафрагму, образованная сферическая волна падает на голограмму и интерферирует с предметной волной.

Рассмотренный интерферометр сходен с ранее представленным трехблочным интерферометром из ФЗП [2], для отображения рентгеновского фазового контраста [3, 4], с той основной отличий, что предметная плоскость перемещена в фокальную плоскость второго блока, а детектор голограммы размещен на месте третьего блока.

Обозначим через $(n, m)$ волновой канал интерферометра, образованный дифракциями $n$- и $m$-порядков соответственно на первом и втором блоках интерферометра. С учетом только 0, –1 и +1 порядков дифракции, в интерферометре останется 9 каналов распространения. Из них (+1, +1) и (0, +1) – предметная и опорная волны соответственно. Рассмотрим поведение остальных 7 каналов. В приближении геометрической оптики, построением траекторий лучей можно показать, что пять из этих каналов – каналы (0, 0), (0, –1), (–1, 0), (–1, +1) и (–1, –1) полностью поглощаются нижней половиной экрана на предметной плоскости, если радиус центральной диафрагмы меньше расстояния ($d$) края ножа от оптической оси (в рассмотренном нами случае радиус диафрагмы выбран $d/2$).

Рассмотрим канал распространения (+1, 0). При условии

$$d > R/3, \qquad (1)$$

где $R$ – радиус ФЗП, канал (+1, 0) не пересекается с голограммой. В случае же несоблюдения этого условия, что предпочтительно с точки зрений увеличения размеров исследуемого образца, часть канала (+1, 0) падает на голограмму. Если интерференция между опорной и предметной волн образует изображение предмета, то интерференции канала (+1, 0) с этими волнами образуют два типа шумов на восстановленном изображении. Шум, образованной в результате интерференции между каналом (+1, 0) и опорной волной обозначим через $n_{\text{ref}}$, а шум, соответствующей интерференции канала (+1, 0) с предметной волной – через $n_{\text{obj}}$.

Рассмотрим местоположение этих, так называемых «интерференционных шумов». На рис.2 приведены траектории лучей канала (+1, 0), опорной и предметной волн, собирающихся в определенной точке (P′) голограммы. Как видно из рисунка, углы между этими лучами связаны соотношениями

$$\varphi_{\text{n,ref}} = (2/3)\varphi_0, \quad \varphi_{\text{n,obj}} = (1/3)\varphi_0, \qquad (2)$$

где $\varphi_0$ – угол между лучами опорной и предметной волн, $\varphi_{\text{n,ref}}$ – между лучами канала (+1, 0) и опорной волны, а $\varphi_{\text{n,obj}}$ – между лучами канала (+1, 0) и предметной волны.

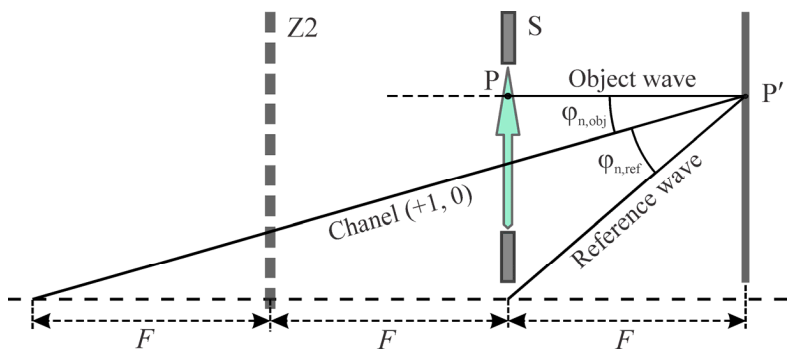

Рис.2. К расчетам местоположений «интерференционных шумов».

Так как восстановление изображения сводится к двумерному обратному Фурье-преобразованию распределения интенсивности голограммы [5], то отображение интерференционной картины на восстановленном изображении удалено от начала координат на расстояние, пропорциональное углу между интерферирующими пучками, в направлении, перпендикулярном к интерференционным полосам. Отсюда, с учетом (2), получим соотношения для $y$-координат отображений интерференционных картин, образованных в окрестности точки P′ голограммы:

$$y_{\text{n,ref}} = (2/3)y_0, \quad y_{\text{n,obj}} = (1/3)y_0. \qquad (3)$$

Здесь $y_{\text{n,ref}}$ и $y_{\text{n,obj}}$ соответствуют шумам $n_{\text{ref}}$ и $n_{\text{obj}}$, соответственно, а $y_0$ – изображению точки P предмета (см. рис.2). При выборе масштаба восстановления 1:1, имеем $y_0 = \tilde{y}_0$, где $\tilde{y}_0$ – $\tilde{y}$-координата точки P′ на голограмме.

Аналогичные оценки для канала (+1, –1), образующего шумы $n'_{\text{ref}}$ и $n'_{\text{obj}}$, приводят к следующим результатом

$$d > R/5, \qquad (1')$$

$$\varphi'_{\text{n,ref}} = 0.6\varphi_0, \quad \varphi'_{\text{n,obj}} = 0.4\varphi_0, \qquad (2')$$

$$y'_{\text{n,ref}} = 0.6y_0, \quad y'_{\text{n,obj}} = 0.4y_0, \qquad (3')$$

где новые обозначения отличаются от старых, прибавлением верхнего штриха. Следует иметь ввиду, что на голограмме интенсивность канала (+1, –1) меньше интенсивности канала (+1, 0).

Интерференция между каналами (+1, 0) и (+1, –1) не рассматривается, так как в этом случае угол интерферирующих лучей слишком мал и соответствующий шум довольно близко расположен к началу координат.

На рис.3 приведен (a) исходный объект и (b) восстановленное из голограммы его изображение. Горизонтальными линиями, на рис.3b, отмечены границы интерференционных шумов при малых значениях $x$, определенных согласно (3) и (3′), с учетом возможного усечения каналов (+1, 0) и (+1 –1), при прохождении через предметную плоскость. Как видно из рисунка, приведенные значения соответствуют восстановленному изображению, за исключением нижней границы шума $n_{\text{ref}}$, что может быть объяснен дифракций канала (+1, 0) на краю ножа K (см. рис.1).

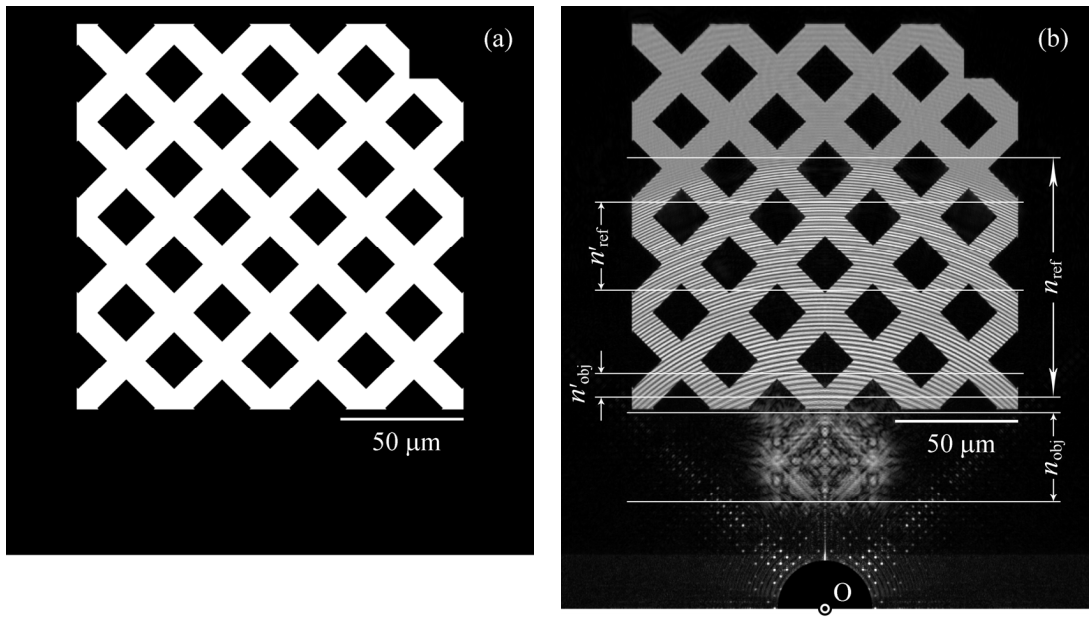

Рис.3. (a) Исследуемый предмет и (b) его изображение, восстановленное из голограммы, с отмеченными областями «интерференционных шумов». Точка O на рис.3b – начало осей координат ($x$, $y$). Внизу восстановленного изображения заметен тень непрозрачного полукруга, расположенного с целью подавления центрального изображения.

Расчеты сделаны для использованных в работе [1] численных значений основных параметров эксперимента; а именно: длина волны рентгеновского излучения – $\lambda$ = 0.1 нм, радиус ФЗП – $R$ = 275.7 мкм, ширина последней зоны Френеля – $\Delta R$ = 181.5 нм, фокальное расстояние первого порядка дифракции – $F$ = 1 м, глубина травления зонной структуры кремниевых ФЗП – $h$ = 9.48 мкм, расстояние края ножа от оптической оси интерферометра – $d$ = 43.1 мкм.

### 3. Подавление шумов восстановленного изображения

В основе рассматриваемого метода подавления шумов лежит обстоятельство, согласно которому «интерференционные шумы» и основные компоненты изображения объекта в определенной точке восстановленного изображении, формируются из разных областей голограммы. Из (3) и (3′), для $\tilde{y}$-координат областей голограммы формирующие шумы $n_{\text{ref}}$, $n_{\text{obj}}$, $n'_{\text{ref}}$, $n'_{\text{obj}}$ и низкочастотные компоненты изображения предмета (обозначим их как $\tilde{y}_{ref}$, $\tilde{y}_{obj}$, $\tilde{y}'_{ref}$, $\tilde{y}'_{obj}$ и $\tilde{y}_0$, соответственно) в точке изображения с $y$-координатой $y_0$, получим

$$\tilde{y}_{ref} = 1.5 y_0, \quad \tilde{y}_{obj} = 3 y_0,$$
$$\tilde{y}'_{ref} = (5/3) y_0, \quad \tilde{y}'_{obj} = 2.5 y_0, \qquad (4)$$
$$\tilde{y}_0 = y_0.$$

Это особенность позволяет, при восстановлении определенного участка изображения, блокировать интерференционные шумы путем затемнения формирующие их участки голограммы, оставляя открытым область – дающий основной вклад в формировании самого изображения предмета.

Таким образом, восстановление изображения предмета проводится не одним шагом, а послойно – с горизонтальными слоями. При каждом шаге затемняется часть голограммы с $\tilde{y} > \alpha y_{min}$ ($1 < \alpha < 1.5$), где $y_{min}$ – нижний края восстанавливаемого слоя. Согласно соотношениям (4), указанным затемнением блокируются «интерференционные шумы» выше линии $y = (\alpha/1.5) y_{min} < y_{min}$ изображения, а, следовательно, и на рассматриваемом слое. Чем уже восстанавливаемая полоса, тем больше расстояние между его верхней границей и затемненной областью, и соответственно, тем меньше высокочастотных пространственных компонент будут потеряны при восстановлении изображения, из-за затемнения части голограммы. Результат такого послойного восстановления для рассмотренного на рис.3 случая, представлен на рис.4. Ширина полос выбрана $\Delta y = 0.05 y_{min}$, $\alpha = 1.4$. Отметим, что представленный многошаговый подход увеличивает лишь процессорное время численных расчетов, оставляя неизменным продолжительность измерений.

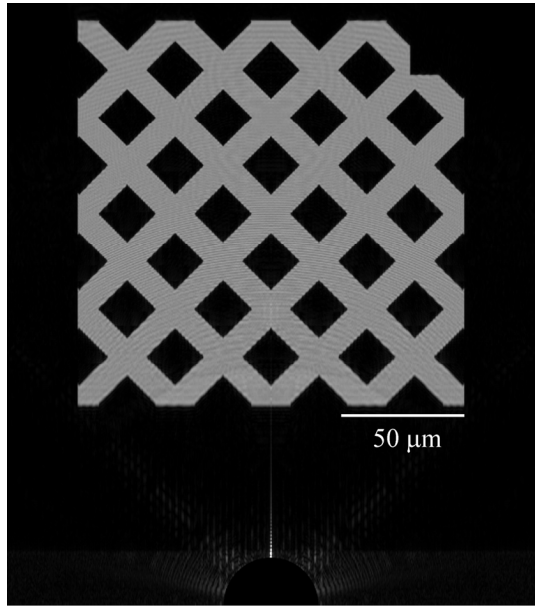

Рис.4. Изображение предмета, восстановленное «послойным» методом.

## 4. Заключение

Работа посвящена исследованию так называемых «интерференционных шумов» возникающих в восстановленном изображении ранее предложенной схемы рентгеновской Фурье-голографии с использованием интерферометра из двух ФЗП с общей оптической осью [1]. Хотя отмеченные шумы исчезают при достаточно большом размере ножа, расположенного перед первым блоком интерферометра (см. рис.1 и условие (1)), такой подход нежелателен из-за малого продольного сечения оставшейся излучения, а, следовательно, и размера исследуемого образца.

Предложена так называемая «послойная» схема восстановления изображения предмета, позволяющий подавлять вышеотмеченные шумы. Недостатком этой схемы является некоторое понижение разрешающей способности восстановленного изображения, что следствие «затемнения» определенных частей голограммы, по ходу послойного восстановления изображения.